\documentclass{svjour2_rev}                    % onecolumn
\smartqed  % flush right qed marks, e.g. at end of proof
\usepackage{graphicx}

\usepackage{amsmath}
\usepackage{amssymb}
\usepackage{bm}
\usepackage{cite}

\begin{document}

\title{
Duality in interacting particle systems and boson representation
}
%\subtitle{Do you have a subtitle?\\ If so, write it here}

%\titlerunning{Short form of title}        % if too long for running head

\author{
Jun Ohkubo
}

%\authorrunning{Short form of author list} % if too long for running head

\institute{
Jun Ohkubo \at
Institute for Solid State Physics, University of Tokyo, \\
Kashiwanoha 5-1-5, Kashiwa, Chiba 277-8581, Japan \\
\email{ohkubo@issp.u-tokyo.ac.jp}
}

%\date{Received: date / Accepted: date}
% The correct dates will be entered by the editor

\maketitle

\begin{abstract}
In the context of Markov processes,
we show a new scheme to derive dual processes and a duality function
based on a boson representation.
This scheme is applicable to a case in which a generator 
is expressed by boson creation and annihilation operators.
For some stochastic processes,
duality relations have been known,
which connect continuous time Markov processes with discrete state space
and those with continuous state space.
We clarify that using a generating function approach and the Doi-Peliti method,
a birth-death process (or discrete random walk model) is naturally connected to 
a differential equation with continuous variables,
which would be interpreted as a dual Markov process.
The key point in the derivation is to use bosonic coherent states
as a bra state, instead of a conventional projection state.
As examples, we apply the scheme to
a simple birth-coagulation process
and a Brownian momentum process.
The generator of the Brownian momentum process is written by elements of the $SU(1,1)$ algebra,
and using a boson realization of $SU(1,1)$
we show that the same scheme is available.
% \PACS{PACS code1 \and PACS code2 \and more}
% \subclass{MSC code1 \and MSC code2 \and more}
\end{abstract}

\section{Introduction}

In the context of nonequilibrium physics,
simple systems of interacting particles have been received considerable attention recently,
and it has been known that the concept of duality is useful
in studying stochastic processes of interacting particle systems \cite{Liggett_book}.
The duality would give deep insights and analytical results for stochastic models in nonequilibrium,
and actually, there are many works using duality properties,
ranging from calculations of correlation functions in interacting particle systems
\cite{Kipnis1982,Spohn1983,Schutz1994,Schutz1997,Doering2003,Giardina2007,Giardina2009} 
to studies of biological population models \cite{Shiga1986,Mohle1999}.

While the duality is a useful concept,
there are some problems to use it.
For example, it has been sometimes necessary to construct a dual process with ad-hoc procedures.
In addition, while the original process and the dual one are connected via a duality function,
the duality function should be selected properly.
To construct the duality function, usually one sets an ansatz for the duality function,
and checks whether the ansatz satisfies the dual relation or not.
Recently, a general procedure to derive a duality function has been proposed \cite{Giardina2009};
dual relations for various stochastic models have been recovered
using the symmetries of the original process.
While this general procedure has wide applications,
it may be needed to use a heuristic way for some specific cases.
For example, 
the general procedure in \cite{Giardina2009} may not be available 
for Brownian momentum (or energy) processes in boundary driven cases,
and a boundary part in a duality function was heuristically found in \cite{Giardina2009}.

In the present paper,
we show a new scheme to obtain a dual process and a duality function.
The scheme is based on a boson representation,
and it is applicable to a case in which a generator of the stochastic process
is expressed by boson creation and annihilation operators.
We will see that a continuous time Markov process with \textit{discrete} state space
(e.g., simple birth-death processes and discrete random walk models)
is a dual process of a continuous time Markov process with \textit{continuous} state space
(e.g., stochastic differential equations and Brownian momentum models).
In the derivation of the duality function,
the generating function approach and the Doi-Peliti method are used.
It will be clarified that 
the dual process and the duality function are naturally derived
by using bosonic coherent states as a bra state in the Doi-Peliti method,
instead of a usual projection state.
In addition, using a boson realization of $SU(1,1)$,
it is also possible to study a duality relation
for a stochastic model with elements of the $SU(1,1)$ algebra.
Especially, we can derive a duality function for a Brownian momentum process with boundaries
not heuristically, but deductively.

The outline of the present paper is as follows.
In section 2, we give a definition of duality.
The new scheme to derive a dual process and a duality function is shown in section 3.
Sections 4 and 5 are applications of the new scheme to two examples;
i.e., a simple birth-coagulation process and a Brownian momentum process.
Section 6 gives concluding remarks.

\section{Duality}

General discussions for duality are given in \cite{Liggett_book}.
In the present paper,
we only treat the duality between birth-death processes (or discrete random walk models) 
and diffusion processes.

Suppose that $(\xi_t)_{t \geq 0}$ and $(z_t)_{t \geq 0}$ are continuous time Markov processes
on state spaces $\Omega$ and $\Omega_\mathrm{dual}$, respectively.
Let $\mathbb{E}_{\xi}$ denotes the expectation
given that the process $(\xi_t)_{t \geq 0}$ starts from $\xi$.
The process $(\xi_t)_{t \geq 0}$ is said to be dual to $(z_t)_{t \geq 0}$
with respect to a duality function $D: \Omega \times \Omega_\mathrm{dual} \to \mathbb{R}$
if for all $\xi \in \Omega$, $z \in \Omega_\mathrm{dual}$ and $t \geq 0$ we have
\begin{align}
\mathbb{E}_{\xi} D(z,\xi_t) = \mathbb{E}_{z}^\mathrm{dual} D(z_t,\xi),
\label{eq_duality}
\end{align}
where $\mathbb{E}_{z}^\mathrm{dual}$ is expectation
in the process $(z_t)_{t \geq 0}$ starting from $z$.

In the following discussions and examples,
the process $(\xi_t)_{t \geq 0}$ is a continuous time Markov process denoting
a birth-death process (or a discrete random walk model),
so that $\xi_t \in \mathbb{N}$.
On the other hand, the dual process $(z_t)_{t \geq 0}$ is a continuous time Markov process
with continuous variables,
and then $z_t \in \mathbb{R}$.

\section{Derivation of duality function using boson representation}

In this section, we derive a duality function.
In order to obtain it, we firstly explain 
a correspondence between a generating function approach and the Doi-Peliti method
(second quantization method).
After that, it will be shown that a duality function is naturally obtained
from a state vector for the continuous time Markov process with discrete state space.

\subsection{Generating function approach}

Some stochastic models with discrete variables are described as birth-death processes.
A time evolution of a birth-death system obeys a master equation,
and it is sometimes useful to treat a generating function 
instead of the original master equation \cite{Gardiner_book}.
For simplicity, we here treat a birth-death process with only one variable.
The generating function $G(x,t)$ is defined as
\begin{align}
G(x,t) = \sum_{n=0}^\infty P(n,t) x^n,
\end{align}
where $n \in \mathbb{N}$, $x \in \mathbb{R}$, 
and $P(n,t)$ is the probability with $n$ particles at time $t$.
The time evolution equation for $G(x,t)$ is written as
\begin{align}
\frac{\mathrm{d}}{\mathrm{d} t} G(x,t) = L\left( x,\frac{\mathrm{d}}{\mathrm{d} x} \right) G(x,t),
\label{eq_time_evol_for_generating_function}
\end{align}
where $L(x,\frac{\mathrm{d}}{\mathrm{d} x})$ is a linear operator,
which is constructed from the original master equation.

\subsection{Doi-Peliti method: boson representation}

The Doi-Peliti method is a well-known approach to investigate birth-death systems
\cite{Doi1976,Doi1976a,Peliti1985}.
In the Doi-Peliti method, bosonic creation and annihilation operators are used:
the creation operator $a^\dagger$ and annihilation operator $a$ satisfy the 
commutation relations
\begin{align}
[a,a^\dagger] = 1, \quad [a, a] = [a^\dagger,a^\dagger] = 0,
\end{align}
and each operator works on a vector in Fock space $| n \rangle$ as follows:
\begin{align}
a^\dagger | n \rangle = | n+1 \rangle, \quad a | n \rangle = n | n-1 \rangle.
\end{align}
The vacuum state $|0\rangle$ is characterized by $a | 0 \rangle = 0$.
The inner product of bra state $\langle m |$ and ket state $| n \rangle$
is defined as
\begin{align}
\langle m | n \rangle = \delta_{m,n} n!,
\end{align}
where $\delta_{m,n}$ is the Kronecker delta.

When we define a time-dependent state $| \psi(t) \rangle$ as
\begin{align}
| \psi(t) \rangle = \sum_{n=0}^{\infty} P(n, t) | n \rangle,
\end{align}
the time evolution of the state $| \phi(t) \rangle$ is given by
\begin{align}
\frac{\mathrm{d}}{\mathrm{d} t} | \psi (t) \rangle = L(a^\dagger,a) | \psi(t) \rangle,
\label{eq_time_evol_for_doi_peliti}
\end{align}
which recovers the original master equation.
The linear operator $L(a^\dagger,a)$ is obtained from the original master equation,
and it is known that
$L(a^\dagger,a)$ in (\ref{eq_time_evol_for_doi_peliti}) has the same form as
$L(x,\frac{\mathrm{d}}{\mathrm{d} x})$ in (\ref{eq_time_evol_for_generating_function}).

While the Doi-Peliti method is similar with usual quantum mechanics,
there are some differences.
One of the big differences is the usage of a projection state.
In the Doi-Peliti method,
the projection state
\begin{align}
\langle \mathcal{P} | \equiv \sum_{n=0}^\infty \frac{1}{n!} \langle n | = \langle 0 | \mathrm{e}^{a}
\end{align}
is used to obtain physical quantities.
For example, 
the average of $n$ is given by
$\sum_{n=0}^\infty n P(n,t) = \langle \mathcal{P} | a^\dagger a | \psi(t) \rangle$.

\subsection{Connection between generating function approach and Doi-Peliti method}

There is a one-to-one correspondence between the generating function approach and the Doi-Peliti method.
Consider the following construction for ket and bra states in the Doi-Peliti method:
\begin{align}
| n \rangle \equiv x^n, \quad 
\langle m | \equiv \int \mathrm{d} x \, \delta(x) \left( \frac{\mathrm{d}}{\mathrm{d} x} \right)^m (\cdot),
\end{align}
and interpret the creation and annihilation operators as follows:
\begin{align}
a^\dagger \equiv x, \quad a \equiv \frac{\mathrm{d}}{\mathrm{d} x}.
\end{align}
Hence, we immediately see that all properties in the Doi-Peliti method are recovered
using $x$ and $\frac{\mathrm{d}}{\mathrm{d} x}$.
In addition, the linear operator $L(a^\dagger,a)$ in (\ref{eq_time_evol_for_doi_peliti})
is obtained by replacing $x$ and $\frac{\mathrm{d}}{\mathrm{d} x}$ of $L(x,\frac{\mathrm{d}}{\mathrm{d} x})$ 
in (\ref{eq_time_evol_for_generating_function}) with $a^\dagger$ and $a$, respectively.

Because of the correspondence between the generating function approach and the Doi-Peliti method,
we will switch between these two notations freely in the following discussions.

\subsection{Derivation of duality function}

For simplicity, a case with only one variable is discussed at first.
After that, a result for general cases will be given.

We consider the following time evolution equation
\begin{align}
\frac{\mathrm{d}}{\mathrm{d} t} | \phi(t) \rangle = L | \phi(t) \rangle,
\end{align}
where $|\phi(t)\rangle$ is given by 
\begin{align}
| \phi(t) \rangle \equiv \sum_{\xi=0}^\infty P(\xi,t) | \xi \rangle, \quad
| \xi \rangle = d( {a^\dagger}, \xi ) | 0 \rangle.
\label{eq_def_of_states}
\end{align}
Here, $\xi$ is a variable used in a continuous time Markov process 
with discrete state space ($\xi \in \mathbb{N}$),
whose probability distribution is denoted by $P(\xi,t)$.
Note that the state $| \xi \rangle$ is not restricted to the form discussed in section 3.2,
and $| \xi \rangle$ is generated using the creation operator $a^\dagger$ via a function $d(a^\dagger,\xi)$.
For example, we will see the following functions $d(a^\dagger,\xi)$ in sections 4 and 5:
\begin{align*}
\textrm{In section 4:} \quad d( {a^\dagger}, \xi ) &= (a^\dagger)^\xi, \\
\textrm{In section 5:} \quad d( {a^\dagger}, \xi ) &= \frac{(a^\dagger)^{2\xi}}{ (2\xi -1)!!},
\end{align*}
where $(2n-1)!! \equiv (2n-1)(2n-3)\cdots 3 \cdot 1$.
In both cases, the state $| \xi \rangle$ is expressed in terms of the creation operator $a^\dagger$,
but it is not necessary to use the simple construction $|n\rangle = (a^\dagger)^n |0\rangle$ in section 3.2.

As explained in section 3.2, 
the projection state is usually used as an adequate `bra' state in the Doi-Peliti formalism.
The key point to obtain a duality function here is the following one;
instead of the projection state,
we define a bra state $\langle \tilde{\phi}(t) |$ as
\begin{align}
\langle \tilde{\phi}(t) | \equiv \int_{-\infty}^\infty \mathrm{d} z \tilde{\phi}(z,t) \langle z |,
\end{align}
where $\langle z |$ is a coherent state of $a^\dagger$:
\begin{align}
\langle z | \equiv \langle 0 | \mathrm{e}^{z a}, 
\end{align}
which satisfies 
\begin{align}
\langle z | a^\dagger = z \langle z |,
\end{align}
and $z$ is assumed to be a real variable.
From the correspondence between the generating function approach and Doi-Peliti method,
the following identities are easily checked:
\begin{align}
\langle z | n \rangle = z^n,
\end{align}
\begin{align}
\langle z | x^k \left( \frac{\mathrm{d}}{\mathrm{d} x} \right)^l | n \rangle
= z^k \left( \frac{\mathrm{d}}{\mathrm{d} z} \right)^l \langle z | n \rangle.
\end{align}
The linear operator $L(a^\dagger,a)$ (i.e, $L(x,\frac{\mathrm{d}}{\mathrm{d} x})$) is generally written in normal order,
i.e., all creation operators are to the left of all annihilation operators in products.
Hence,
\begin{align}
\langle \tilde{\phi}(t) | L \left( x, \frac{\mathrm{d}}{\mathrm{d} x}\right) 
&=\int_{-\infty}^\infty \mathrm{d} z\tilde{\phi}(z,t) \langle z | L \left( x, \frac{\mathrm{d}}{\mathrm{d} x}\right) \nonumber \\
&=\int_{-\infty}^\infty \mathrm{d} z\tilde{\phi}(z,t) L \left( z, \frac{\mathrm{d}}{\mathrm{d} z}\right) \langle z |  \nonumber \\
&= \int_{-\infty}^\infty \mathrm{d} z 
\left[ L^{*} \left( z, \frac{\mathrm{d}}{\mathrm{d} z}\right) \tilde{\phi}(z,t)  \right] \langle z | ,
\label{eq_time_evol_tilde_phi}
\end{align}
where $L\left( z,\frac{\mathrm{d}}{\mathrm{d} z} \right)$ is obtained by simply replacing $x$ and $\frac{\mathrm{d}}{\mathrm{d} x}$
as $z$ and $\frac{\mathrm{d}}{\mathrm{d} z}$, respectively;
$L^{*} \left( z,\frac{\mathrm{d}}{\mathrm{d} z} \right)$ is 
the adjoint operator of $L\left( z,\frac{\mathrm{d}}{\mathrm{d} z} \right)$.
Therefore, we obtain the following identity:
\begin{align}
\langle \tilde{\phi}(0) | \phi(t) \rangle
= \langle \tilde{\phi}(0) | \mathrm{e}^{Lt} | \phi(0) \rangle
= \langle \tilde{\phi}(t) | \phi(0) \rangle,
\label{eq_duality_pre}
\end{align}
where the time development of the bra state $\langle \tilde{\phi}(t) |$ is defined as
\begin{align}
\frac{\mathrm{d}}{\mathrm{d} t} \langle \tilde{\phi}(z,t) | = \langle \tilde{\phi}(z,t) | L.
\label{eq_time_evol_bra}
\end{align}
Combining (\ref{eq_time_evol_tilde_phi}) and (\ref{eq_time_evol_bra}), we have
\begin{align}
\frac{\mathrm{d}}{\mathrm{d} t} \tilde{\phi}(z,t) = L^{*}\left(z,\frac{\mathrm{d}}{\mathrm{d} z} \right) \tilde{\phi}(z,t).
\end{align}
At this stage, it is clarified that 
the linear operator $L$ is a generator for the continuous time Markov process $(z_t)_{t \geq 0}$
if $\tilde{\phi}(z,t)$ can be considered as the time-dependent probability density.
Hence, a continuous time Markov process with discrete state space, $(\xi_t)_{t \geq 0}$,
is naturally connected to the stochastic process with continuous variables $(z_t)_{t \geq 0}$.
In addition, writing (\ref{eq_duality_pre}) explicitly, we have
\begin{align}
\int_{-\infty}^\infty \mathrm{d} z \sum_{\xi=0}^\infty \tilde{\phi}(z,0) P(\xi,t) d(z,\xi)
= \int_{-\infty}^\infty \mathrm{d} z \sum_{\xi=0}^\infty \tilde{\phi}(z,t) P(\xi,0) d(z,\xi).
\end{align}
Hence, if we set the initial conditions for $\xi$ and $z$ as 
a Kronecker delta function and a Dirac delta function respectively,
the duality relation (\ref{eq_duality}) is obtained.
It is also clear that the function $d$ gives a duality function.
The above discussion means that if a generator $L$ is expressed 
in terms of creation and annihilation operators,
a dual process is immediately constructed
and the duality function is given by the function $d$,
which specifies a state $|\xi\rangle$ in the Markov process with discrete state space
(see (\ref{eq_def_of_states})).

If a Markov process with discrete state space has many variables $\{\xi_i\}$ ($i \in \{ 1, \dots, N\}$),
a state $| \xi \rangle$ is defined by 
\begin{align}
| \xi \rangle = \bigotimes_{i=1}^{N} | \xi_i \rangle_i 
= \left( \prod_{i=1}^{N} d_i(a^\dagger_i, \xi_i) \right) 
\left( \bigotimes_{i=1}^{N} | 0 \rangle_i \right),
\end{align}
where $d_i(a_i^\dagger,\xi)$ may be different from each other.
Hence, using 
$z = \{ z_1, z_2, \dots, z_N \}$, $z_i \in \mathbb{R}$, 
and $\xi = \{ \xi_1, \xi_2, \dots, \xi_N \}$, $\xi_i \in \mathbb{N}$,
a duality function is given as
\begin{align}
D(z,\xi) \equiv \prod_{i=1}^{N}  d_i(z_i, \xi_i).
\end{align}

\section{Example 1: Simple birth-coagulation process}

As a first example,
we apply the scheme in section 3 to a simple birth-coagulation process.
The birth-coagulation process has been used widely to study
front-propagation problems,
and it has been known that
a Langevin equation, so-called stochastic Fisher and
Kolmogorov-Petrovsky-Piscounov (sFKPP) equation,
plays an important role in the study of the front-propagation problems
 \cite{Brunet1997,Pechenik1999,Panja2004,Brunet2006}.
Recently, the sFKPP equation has been discussed even in a QCD context \cite{Munier2006}.

A duality relation for the birth-coagulation process
has been used to study a front propagating problem in \cite{Doering2003}.
In the duality relation, the birth-coagulation process is connected to a Langevin equation.
In \cite{Doering2003}, the dual process and the duality function were assumed,
and explicit calculations based on stochastic differential equations were used
to check the duality relation.
We will show that the dual process and the duality function are recovered simply
using our general scheme.
In addition, we will derive a new duality relation for a slightly-changed stochastic process.
The derivation demonstrates the effectiveness of the present scheme
to find a new duality relation.

\subsection{Derivation of duality relation in the birth-coagulation process}
\label{sec_ex1_1}

Consider the following reaction scheme for the birth-coagulation process:
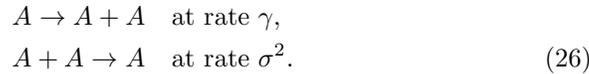
\begin{align}
&A \to A+A \quad \textrm{at rate $\gamma$}, \nonumber \\
&A + A \to A \quad \textrm{at rate $\sigma^2$}.
\label{eq_ex1_process}
\end{align}
The master equation for the birth-coagulation process is written as
\begin{align}
\frac{\mathrm{d}}{\mathrm{d} t} P(\xi,t) 
=&
\gamma (\xi-1) P(\xi-1,t) - \gamma \xi P(\xi,t) \nonumber \\
&- \sigma^2 \frac{\xi(\xi-1)}{2}P(\xi,t) + \sigma^2 \frac{(\xi+1)\xi}{2} P(\xi+1,t),
\label{eq_ex1_master_equation}
\end{align}
where $\xi$ is the number of particle $A$, and $\xi \in \mathbb{N}$.
The linear operator in the Doi-Peliti method is given by
\begin{align}
L = \gamma(a^\dagger-1) a^\dagger a + \frac{\sigma^2}{2}(1-a^\dagger) a^\dagger a^2,
\label{eq_generator_ex1}
\end{align}
and it is easy to check that the following time evolution equation and a bra state $| \phi(t)\rangle$
recover the master equation (\ref{eq_ex1_master_equation}):
\begin{align}
\frac{\mathrm{d}}{\mathrm{d} t} | \phi(t) \rangle = L | \phi(t) \rangle, \quad
| \phi(t) \rangle = \sum_{\xi=0}^\infty P(\xi,t) | \xi \rangle
\end{align}
and 
\begin{align}
| \xi \rangle = (a^\dagger)^{\xi} | 0 \rangle.
\end{align}
Since $d(a^\dagger,\xi) = (a^\dagger)^{\xi}$,
a duality function is 
\begin{align}
D(z,\xi) = z^\xi.
\label{eq_ex1_duality_function}
\end{align}
Considering the adjoint of the linear operator $L$ in terms of $z$ and $\frac{\mathrm{d}}{\mathrm{d} z}$,
\begin{align}
L^{*} = - \frac{\mathrm{d}}{\mathrm{d} z} \left[ -\gamma z (1-z) \right]
+ \frac{1}{2} \frac{\mathrm{d}^2}{\mathrm{d} z^2} \left[ \sigma^2 (1-z) z \right],
\end{align}
we see that the adjoint operator $L^{*}$ gives a Fokker-Planck equation.
Hence, the dual process corresponds to the following stochastic differential equation:
\begin{align}
\mathrm{d} z = - \gamma z(1-z) \mathrm{d} t + \sigma \sqrt{z(1-z)} \mathrm{d} W.
\end{align}
If one consider a new process via a variable transformation $u(t) = 1-z(t)$,
the corresponding stochastic differential equation is
\begin{align}
\mathrm{d} u = \gamma u(1-u) \mathrm{d} t + \sigma \sqrt{u(1-u)} \mathrm{d} W,
\end{align}
and the duality function is rewritten by using the new variable $u$ as
\begin{align}
D(u,\xi) = (1-u)^\xi.
\end{align}
The above dual process and the duality function are consistent with results in \cite{Doering2003}.

\subsection{New duality relation for a slightly-changed reaction scheme}

We here show that it is easy to derive a new duality relation for a slightly-changed reaction scheme,
which has not been studied yet.

We consider that the following reaction is added to the stochastic system (\ref{eq_ex1_process}):
\begin{align}
A \to A + A + A \quad \textrm{at rate $\alpha$}.
\end{align}
Hence, a new term,
$\alpha (\xi -2) P(\xi-2,t) - \alpha \xi P(\xi,t)$,
is added to the master equation (\ref{eq_ex1_master_equation}).
The corresponding linear operator in the Doi-Peliti method is
$\alpha ( (a^\dagger)^2 -1) a^\dagger a$,
and then the adjoint operator $L^*$ is finally given by
\begin{align}
L^{*} = - \frac{\mathrm{d}}{\mathrm{d} z} \left[ -\gamma z (1-z) 
- \alpha z(1-z^2)
\right]
+ \frac{1}{2} \frac{\mathrm{d}^2}{\mathrm{d} z^2} \left[ \sigma^2 (1-z) z \right].
\end{align}
Thus, we conclude that 
the dual process is given by the following stochastic differential equation:
\begin{align}
\mathrm{d} z = 
- [\gamma z(1-z) + \alpha z(1-z^2) ]\mathrm{d} t + \sigma \sqrt{z(1-z)} \mathrm{d} W,
\end{align}
and the duality function is given by (\ref{eq_ex1_duality_function}).

\section{Example 2: Brownian momentum process}

As a next example, a Brownian momentum process is studied \cite{Giardina2007,Giardina2009}.
In this case, different from example 1 in section 4,
we start from a stochastic process with continuous state space
and obtain a dual process with discrete state space.

The model is defined as a stochastic process 
on $N$-dimensional vectors $(z_1, \dots, z_N) \in \mathbb{R}^N$,
which have to be interpreted as momenta associated with lattice sites $\{1, \dots, N \}$.
In addition, sites $1$ and $N$ are in contact with heat reservoirs
at temperature $T_\mathrm{L}$ and $T_\mathrm{R}$, respectively.
The process is defined by a generator $L$ as follows:
\begin{align}
L = L_1 + L_N + \sum_{i=1}^{N-1} L_{i,i+1},
\end{align}
with
\begin{align}
L_1 f &= T_\mathrm{L} \frac{\partial^2}{\partial z_1^2} f - z_1 \frac{\partial}{\partial z_1} f,\\
L_N f &= T_\mathrm{R} \frac{\partial^2}{\partial z_N^2} f - z_N \frac{\partial}{\partial z_N} f,\\
L_{i,i+1} f &= \left( z_{i} \frac{\partial}{\partial z_{i+1}} -
z_{i+1} \frac{\partial}{\partial z_{i}}  \right)^2 (f),
\end{align}
where $f$ is a $\mathcal{C}^\infty$ function.
Hence, the time-dependent probability density $p(z,t)$ obeys the following equation:
\begin{align}
&\frac{\partial}{\partial t} p(z,t) = L^{*} p(z,t),\\
&L^{*} = L^{*}_1 + L^{*}_N + \sum_{i=1}^{N-1} L^{*}_{i,i+1},
\end{align}
where
\begin{align}
L^{*}_1 f &= T_\mathrm{L} \frac{\partial^2}{\partial z_1^2} f + \frac{\partial}{\partial z_1} (z_1 f),\\
L^{*}_N f &= T_\mathrm{R} \frac{\partial^2}{\partial z_N^2} f + \frac{\partial}{\partial z_N} (z_N f),\\
L^{*}_{i,i+1} &= L_{i,i+1}.
\end{align}
The above equation corresponds to a stochastic process with continuous state space \cite{Giardina2007}.

It has been known that the generator $L$ is rewritten 
by using elements (operators) of the $SU(1,1)$ algebra \cite{Giardina2009}.
The operators are defined by
\begin{align}
K_i^{+} = \frac{1}{2} z_i^2, \quad
K_i^{-} = \frac{1}{2} \frac{\partial^2}{\partial z_i^2}, \quad
K_i^{0} = \frac{1}{4} \left(\frac{\partial}{\partial z_i}(z_i \, \cdot) 
+ z_i \frac{\partial}{\partial z_i} \right),
\label{eq_def_of_su11}
\end{align}
and they satisfy the following commutation relations:
\begin{align}
[K_i^0,K_i^{\pm}] = \pm K_i^{\pm}, \quad [K_i^{-}, K_i^{+}] = 2 K_i^0.
\end{align}
The components of the generator $L$ is rewritten as
\begin{align}
L_{i,i+1} &= 4 \left(
K_i^{+} K_{i+1}^{-} + K_i^{-} K_{i+1}^{+} - 2 K_i^{0} K_{i+1}^{0} + \frac{1}{8}
\right), \\
L_1 &= 2 T_\mathrm{L} K_1^{-} - 2 K_1^{0} + \frac{1}{2},\\
L_N &= 2 T_\mathrm{R} K_N^{-} - 2 K_N^{0} + \frac{1}{2}.
\end{align}
In addition, the $SU(1,1)$ group admits a discrete (infinite dimensional) representation:
\begin{align}
K_i^{+} | \xi_i \rangle = \left( \frac{1}{2} + \xi_i \right) | \xi_i + 1 \rangle, \quad
K_i^{-} | \xi_i \rangle = \xi_i | \xi_i - 1 \rangle, \quad
K_i^{0} | \xi_i \rangle = \left( \xi_i + \frac{1}{4} \right) | \xi_i  \rangle.
\label{eq_discrete_representation_of_su11}
\end{align}
Here, we reinterpret the operators $L_1$ and $L_N$ as follows:
\begin{align}
L_1 &= 2 T_\mathrm{L} K_1^{-} - 2 K_1^{0} + \frac{1}{2}
\equiv 2 a^\dagger_0 K_1^{-} - 2 K_1^{0} + \frac{1}{2}, \\
L_N &= 2 T_\mathrm{R} K_N^{-} - 2 K_N^{0} + \frac{1}{2}
\equiv 2 a^\dagger_{N+1} K_N^{-} - 2 K_N^{0} + \frac{1}{2},
\end{align}
where we interpret the constants $T_\mathrm{L}$ and $T_\mathrm{R}$
as the creation operators related to additional sites $0$ and $N+1$;
$T_\mathrm{L} = z_0 \equiv a^\dagger_0$ and $T_\mathrm{R} = z_{N+1} \equiv a^\dagger_{N+1}$.
This reinterpretation is justified according to the correspondence
between the generating function approach and the Doi-Peliti method (see section 3.3).
Using the introduction of 
the bosonic creation operators for sites $0$ and $N+1$
and the $SU(1,1)$ algebra for the other sites,
it is possible to consider that the generator $L$
creates a stochastic process with discrete state space.
We define, for $\xi \in \Omega$, $i,j \in \{0,\dots,N+1\}$,
the configuration $\xi^{i,j}$ to be the configuration obtained from $\xi$
by removing one particle at $i$ and adding one particle at $j$.
Hence, the linear operator $L$ is interpreted as 
\begin{align}
&L \psi(\xi) = \nonumber \\
&2 \xi_1 [ \psi( \xi^{1,0}) - \psi(\xi) ] + 2 \xi_1 (2\xi_2 + 1) [ \psi(\xi^{1,2})-\psi(\xi)] \nonumber \\
&+ \sum_{i=2}^{N-1} \Big(
2\xi_i (2\xi_{i-1}+1)[ \psi(\xi^{i,i-1}) - \psi(\xi) ] 
+ 2\xi_i (2\xi_{i+1}+1)[ \psi(\xi^{i,i+1}) - \psi(\xi) ]
\Big) \nonumber \\
&+ 2 \xi_N (2\xi_{N-1} + 1) [ \psi(\xi^{N,N-1})-\psi(\xi)] + 2 \xi_N [ \psi( \xi^{N,N+1}) - \psi(\xi) ] ,
\label{eq_generator_ex2}
\end{align}
where $\psi: \Omega \to \mathbb{R}$ is an arbitrary function of the finite particle configurations.
This process is considered as a discrete random walk model 
with absorbing sites $0$ and $N+1$\cite{Giardina2007}.
Hence, a stochastic process with continuous variables (the Brownian momentum process)
is naturally connected to a stochastic process with discrete variables (discrete random walk model).

Next, we obtain the duality function.
Using (\ref{eq_discrete_representation_of_su11}) iteratively,
it is easy to confirm that a state $| \xi_i \rangle$ for $i \in \{ 1, \dots, N \}$ is given by
\begin{align}
| \xi_i \rangle = \frac{2^{\xi_i}}{(2\xi_i - 1)!!} (K_i^{+})^{\xi_i} | 0 \rangle_i',
\end{align}
where $| 0  \rangle_i'$ is the vacuum state in the $SU(1,1)$ representation.
Here, we note that 
there are boson representations for the operators in $SU(1,1)$  \cite{Perelomov_book}:
\begin{align}
K_i^{+} = \frac{1}{2} (a^\dagger_i)^2, \quad
K_i^{-} = \frac{1}{2} (a_i)^2, \quad
K_i^{0} = \frac{1}{4} (a_i a_i^\dagger + a^\dagger_i a_i).
\end{align}
(See also (\ref{eq_def_of_su11}).)
Hence, it is possible to rewrite state $| \xi_i \rangle$ by using boson creation operators
instead of $K_i^{+}$.
On the other hand, for site $0$ and $N+1$,
it is necessary to indicate a state $| \xi_0 \rangle_0$ ($| \xi_{N+1} \rangle_{N+1}$)
by using the creation operator $a^\dagger_0$ ($a^\dagger_{N+1}$) instead of $K_0^{+}$ ($K_{N+1}^{+}$);
i.e., $| \xi_0 \rangle_0 = (a^\dagger_0)^{\xi_0} | 0 \rangle_0$ 
and $| \xi_{N+1} \rangle_{N+1} = (a^\dagger_{N+1})^{\xi_{N+1}} | 0 \rangle_{N+1}$.
We therefore obtain
\begin{align}
| \xi \rangle &= | \xi_0 \rangle_0 \left( \bigotimes_{i=1}^{N} | \xi_i \rangle_i \right)
\otimes | \xi_{N+1} \rangle_{N+1} \nonumber \\
&= \left( (a^\dagger_0)^{\xi_0} | 0 \rangle_0 \right) 
\left( \bigotimes_{i=1}^N \left[ \frac{1}{(2\xi_i-1)!!} (a^\dagger_i)^{2\xi_i} | 0 \rangle_i \right] \right)
\otimes \left(  (a^\dagger_{N+1})^{\xi_{N+1}} | 0 \rangle_{N+1} \right) .
\end{align}
This means 
\begin{align}
d_i(a_i^\dagger,\xi_i) =
\begin{cases}
(a_i^\dagger)^{\xi_i} & \textrm{for } i = 0, N+1, \\
\frac{(a_i^\dagger)^{2\xi_i}}{ (2\xi_i -1)!!} & \textrm{otherwise},
\end{cases}
\end{align}
and then we immediately have a duality function as
(note that $z_0 = T_L$ and $z_{N+1} = T_R$)
\begin{align}
D(z,\xi) = T_L^{\xi_0} T_R^{\xi_{N+1}} \prod_{i=1}^N \frac{z_i^{2\xi_i}}{(2\xi_i - 1)!!},
\end{align}
which is consistent with the results of \cite{Giardina2007,Giardina2009}.

\section{Concluding remarks}

In the present paper, a new scheme to obtain a dual process and duality function
was given; the scheme is applicable when a generator is expressed
in terms of boson representation.
Using the new scheme, it is possible to find the duality function
not heuristically, but deductively using a state expression 
for a continuous time Markov process with discrete state space.
The duality connects the Markov process to the dual process with continuous state space.
We applied the scheme to two examples, and adequately recovered results in previous works,
and derived a new duality relation for a slightly-changed stochastic process.
We here note that the same scheme is available to obtain a duality function
in a Brownian energy model in \cite{Giardina2009},
and we confirmed that a suitable duality function is actually obtained.

Our analysis in the present paper was limited to simple cases
in which a representation of the $x$-representation is changed to the $n$-representation.
In \cite{Giardina2009}, the role of symmetries was considered,
which then leads to self-duality for discrete processes.
The key idea in our scheme is a usage of coherent states as a bra state,
instead of a conventional projection state.
Although we here limited ourselves to cases with boson representation, 
the similar idea will be also applicable to other cases.
For example, it will be possible to apply the similar discussions
to a case with a generator with the $SU(2)$ algebra (or the quantum algebra $U_q[SU(2)]$),
which describes symmetric (or asymmetric) simple exclusion processes \cite{Schutz1997}.
Such extensions are out of the scope of the present paper;
these issues are currently under investigations and will be published in future.

Finally, we comment on the applicability of the scheme.
The scheme would basically connect a discrete representation and continuous one,
and give a duality function between them.
However, there may be no guarantee that
a corresponding dual `process' adequately describes a stochastic process;
e.g., it is necessary for Markov processes with discrete state space
to have the generators with some specific forms
(see (\ref{eq_generator_ex1}) and (\ref{eq_generator_ex2})).
In our scheme,
it is at least possible to derive a differential equation with continuous variables,
which may not be interpreted as a stochastic process.
The derived differential equation and the original Markov process
is adequately connected via a (duality) function,
which is simply given from a state expression for the Markov process, 
as discussed in the present paper.
Such `duality' between a stochastic process and a deterministic differential equation
would be valuable for studies of nonequilibrium physics.
In addition, the present formalism based on the generating function and Doi-Peliti method
would be tractable for physicists,
and then it will help to seek new duality relations for stochastic processes.

\begin{acknowledgements}
The author thank S. Sasa and K. Itakura for motive argument of this work.
This work was supported in part by grant-in-aid for scientific research 
(Grants No.~20115009 and No.~21740283)
from the Ministry of Education, Culture, Sports, Science and Technology, Japan.
\end{acknowledgements}

% BibTeX users please use one of
%\bibliographystyle{spbasic}      % basic style, author-year citations
%\bibliographystyle{spmpsci}      % mathematics and physical sciences
%\bibliographystyle{spphys}       % APS-like style for physics
%\bibliography{}   % name your BibTeX data base

\begin{thebibliography}{99}
% Liggett book
\bibitem{Liggett_book} Liggett T M 2005 
{\it Interacting Particle Systems (Classics in Mathematics)} (Berlin: Springer)
Reprint of the 1985 edition

% application of duality
\bibitem{Kipnis1982} Kipnis C, Marchioro C and Presutti E 1982 {\it J. Stat. Phys.} {\bf 27} 65
\bibitem{Spohn1983} Spohn H 1983 {\it J. Phys. A: Math. Gen.} {\bf 16} 4275
\bibitem{Schutz1994} Sch{\"u}tz G and Sandow S 1994 {\it Phys. Rev. E} {\bf 49} 2726
\bibitem{Schutz1997} Sch{\"u}tz G M 1997 {\it J. Stat. Phys.} {\bf 86} 1265
\bibitem{Doering2003} Doering C R, Mueller C and Smereka P 2003 {\it Physica A} {\bf 325} 243
% duality in Brownian momentum process
\bibitem{Giardina2007} Giardin{\`a} C, Kurchan J and Redig F 2007 {\it J. Math. Phys.} {\bf 48} 033301
\bibitem{Giardina2009} Giardin{\`a} C, Kurchan J, Redig F and Vafayi K 
2009 {\it J. Stat. Phys.} {\bf 135} 25

% duality in population model
\bibitem{Shiga1986} Shiga T and Uchiyama K 1986 {\it Probab. Th. Rel. Fields} {\bf 73} 87
\bibitem{Mohle1999} M{\"o}hle M 1999 {\it Bernoulli} {\bf 5} 761

% generating function and Fokker-Planck
\bibitem{Gardiner_book} Gardiner C W 2004 {\it Handbook of Stochastic Methods} 3rd edn (Berlin: Springer)


% Doi-Peliti
\bibitem{Doi1976} Doi M 1976 {\it J. Phys. A: Math. Gen.} {\bf 9} 1465
\bibitem{Doi1976a} Doi M 1976 {\it J. Phys. A: Math. Gen.} {\bf 9} 1479
\bibitem{Peliti1985} Peliti L 1985 {\it J. Physique} {\bf 46} 1469


% front (sFKPP)
\bibitem{Brunet1997} Brunet E and Derrida B 1997 {\it Phys. Rev. E} {\bf 56} 2597
% front (sFKPP and Doi-Peliti)
\bibitem{Pechenik1999} Pechenik L and Levine H 1999 {\it Phys. Rev. E} {\bf 59} 3893
% effects of fluctuations on propagating fronts (review)
\bibitem{Panja2004} Panja D 2004 {\it Phys. Rept.} {\bf 393} 87
% front (sFKPP)
\bibitem{Brunet2006} Brunet E, Derrida B, Mueller A H and Munier S 2006 {\it Phys. Rev. E} {\bf 73} 056126
% birth-coagulation process (sFKPP) in QCD
\bibitem{Munier2006} Munier S 2006 {\it Acta Phys. Polo. B} {\bf 37} 3451


% boson realization of SU(1,1)
\bibitem{Perelomov_book} Perelomov A 1986 
{\it Generalized Coherent States and Their Applications} (Berlin: Springer)
\end{thebibliography}

% Non-BibTeX users please use

\end{document}